\documentstyle[prb,aps,epsfig,multicol]{revtex}

\begin{document}
\draft

\title{Meissner - London state in superconductors of rectangular cross-section
in perpendicular magnetic field}

\author{R. Prozorov and R. W. Giannetta}
\address{Loomis Laboratory of Physics, University of Illinois at Urbana-
Champaign,\\ 1110 West Green Street, Urbana, Illinois 61801}

\author{A. Carrington}
\address{Department of Physics and Astronomy, University of Leeds, Leeds LS2
9JT, United Kingdom}

\author{F. M. Araujo-Moreira}

\address{Grupo de Supercondutividade e Magnetismo, Departamento de
F$\acute{\imath}$sica,
Universidade Federal de S$\tilde{a}$o Carlos, Caixa Postal 676,
S$\tilde{a}$o Carlos SP, 13565-905 Brazil.}

\date{submitted to Phys. Rev. B, February 29, 2000}
\maketitle

\begin{abstract}
The distribution of magnetic induction in Meissner state with finite
London penetration depth is analyzed for platelet samples of rectangular
cross-section in a perpendicular magnetic field. The exact 2D numerical
solution of the London equation is extended analytically to the realistic
3D case. Data obtained on Nb cylinders and foils as well as single
crystals of YBCO and BSCCO are in a good agreement with the
model. The results are particularly relevant for magnetic susceptibility,
rf and microwave resonator measurements of the magnetic penetration depth
in high-$T_{c}$ superconductors.
\end{abstract}


\begin{multicols}{2}
\narrowtext

The temperature and field dependencies of the magnetic penetration
depth yield basic information about the microscopic pairing state
of a superconductor \cite{dwave} as well as vortex static and
dynamic behavior \cite{brandt2,clem}. Since most high-$T_{c}$
superconductors are highly anisotropic, a measurement in which the
applied magnetic field lies at an arbitrary angle relative to the
conducting planes yields a Meissner response arising from both
in-plane and inter-plane supercurrents. The corresponding
penetration depths $\lambda_{ab}$ and $\lambda_{c}$ can differ
widely in their magnitude and temperature dependence and it is
desirable to separate the two contributions to the measured
penetration depth. To study $\lambda_{ab}$ one must resort to a
configuration in which the applied field is normal to the
conducting planes so as to generate only in-plane supercurrents.
Unfortunately, the London equations in this geometry cannot be
solved analytically, making it difficult to reliably relate the
experimental response (typically a frequency shift or change in
magnetic susceptibility) to changes in $\lambda_{ab}$. Exact
analytical solutions are known only for special geometries: an
infinite bar or cylinder in longitudinal field, a cylinder in
perpendicular field, a sphere, or a thin film. These solutions are
not practical since most high-$T_{c}$ superconducting crystals are
thin plates with aspect ratios typically ranging from 1 to 30.
Brandt developed a general numerical method to calculate magnetic
susceptibility for plates and discs \cite{brandt2} but this method
is difficult to apply in practice and the solutions are limited to
two dimensions.

In this paper we describe the numerical solution of the London equations
in two dimensions for long slabs in a perpendicular field. The results
are then extended analytically to three dimensions.  We first compare our
calculations in the limit of $\lambda=0$ with SQUID measurements on
cylindrical Nb samples of differing aspect ratio \cite{fernando}. We then
compare our calculations for finite $\lambda$ with data from Nb foils and
platelets of both BSCCO and YBCO high-$T_{c}$ superconductors, obtained
by using an rf LC resonator \cite{resonator}. Using numerical results and
analytical approximations we derive a formula which can be used to
interpret frequency shift data obtained from rf and microwave resonator
experiments as well as sensitive magnetic susceptibility measurements.
\begin{figure}
 \centerline{\psfig{figure=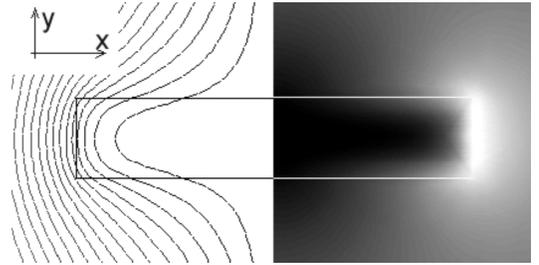,width=7cm,clip}}
 \caption{{\it Right half}: gray scale image of the magnetic field in and around
the sample of $d/w=1/5$ and $\lambda/d=0.5$. Black color represents
$B=0$. {\it Left half}: contour lines of the vector potential.}
 \label{gray}
\end{figure}

Consider an isotropic superconducting slab of width $2w$ in the
$x$-direction, thickness $2d$ in the $y$-direction, and infinite in $z-$
direction. A uniform magnetic field $H_{0}$ is applied along the $y$ -
direction. In this $2D$ geometry the vector potential is ${\bf
A}=\{0,0,A\}$, so that the magnetic field has only two components ${\bf
H}=\{\partial A/\partial y,-\partial A/\partial x,0\}$ and the London
equation takes the form: $\Delta A-\lambda ^{-2} A=0$. Outside the sample
$\Delta A=-4 \pi j/c=0$ and $\partial A/\partial n$ is continuous along
the sample boundary. Here $n$ is the direction normal to the sample
surface. A numerical solution of this equation was obtained using the
finite-element method on a triangular adaptive mesh using Gauss-Newton
iterations scheme. The boundary conditions were chosen to obtain constant
magnetic field far from the sample, i.e., $A\left( x,y\right) =-H_{0}x$
for $y>>d$ and $x$ $>>w$.

\begin{figure}
 \centerline{\psfig{figure=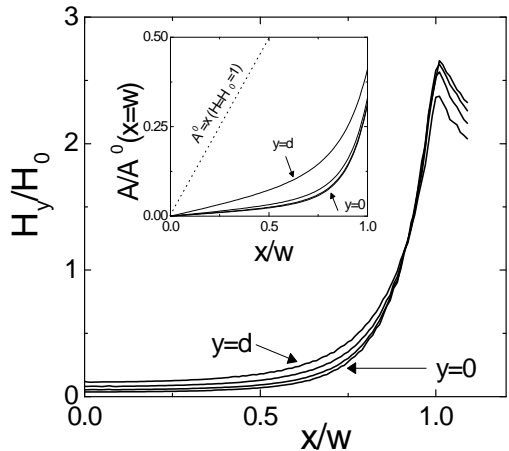,width=7cm,clip}}
 \caption{Profiles of the y-component of the magnetic field (parallel to the
external field) for the sample shown in Fig. \ref{gray}.
 {\it Inset} corresponding profiles of the vector potential.}
 \label{prof}
\end{figure}

Figure \ref{gray} presents the distribution of the the magnetic field in
and around the sample with $w/d=5$ and $\lambda /d=0.5$. The black color
on a gray scale image corresponds to $\left| {\bf B} \right|=0$. The left
half of the sample shows contour lines of the vector potential. Figure
\ref{prof} shows profiles of the y-component of the magnetic field at
different distances $y$ from the sample middle plane.

\begin{figure}
 \centerline{\psfig{figure=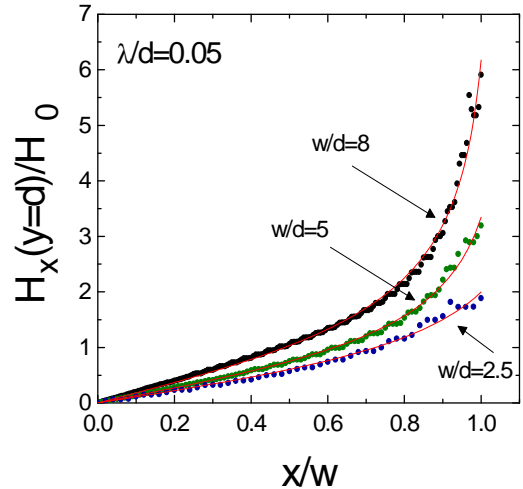,width=7cm,clip}}
 \caption{Distribution of the in-plane, $H_{x}$, component of the magnetic
field on the sample surface, $y=d$. Symbols show result of numerical
calculation and solid lines are the fits to Eq.(\ref{hs}).}
 \label{hsurface}
\end{figure}

The inset shows the corresponding profiles of the vector potential,
normalized by its value $A^{0}(x=w)$ in the absence of a sample (a
uniform-field curve $A^{0}=x$ is shown by the dotted line). Using the
London relation $4\pi\lambda ^{2}j=-cA$ and the definition of the
magnetic moment $M=(2c)^{-1}\int {\bf r} \times {\bf j}d^{3}r$ we
calculate numerically the susceptibility per unit volume (unit of surface
cross-section in 2D case):
\begin{equation}
4 \pi \chi =\frac{1}{dw\lambda
^{2}H_{0}}\int\limits_{0}^{d}dy\int\limits_{0}^{w}A%
\left( x,y\right) xdx
\label{chi}
\end{equation}
It is easy to check that for an infinite slab of width $2w$ in parallel
field, where $A=-\lambda H_{0} \sinh{(x/\lambda)}/\cosh{(w/\lambda)}$,
Eq.(\ref{chi}) results in a known expression similar to Eq.(\ref{chifit})
below (with $N=0$ and $R=w$). In finite geometry there will be a
contribution to the total susceptibility from the currents flowing on top
and bottom surfaces. These currents are due to shielding of the in-plane
component of the magnetic field, $H_{x}=\partial A/\partial y$, appearing
due to demagnetization. Figure \ref{hsurface} shows profiles of $H_{x}$ on
the sample surface, at $y=d$, calculated for three different samples,
$w/d=$ 8, 5, and 2.5.  An analytical form for the surface magnetic field is
known only for elliptical samples. We find, however, that it can be mapped
onto the flat surface, so that the distribution of $H_{x}$ is given by:
\begin{equation}
H_{x} =\frac{H_{0} r}{\sqrt{a^2-r^2}}
\label{hs}
\end{equation}
\noindent where $r \equiv x/w$ and $a^{2} = 1 + (2d/w)^{2}$. This equation
is similar to that obtained for an ideal Meissner screening
\cite{norris,fabbricatore}. Solid lines in Fig.\ref{hsurface} are the fits
to Eq.(\ref{hs}). The agreement between numerical and analytical results
is apparent.

Next we find a simple analytical approximation to the exact numerical
results by calculating the ratio of the volume penetrated by the magnetic
field to the total sample volume. This procedure automatically takes into
account demagnetization and non-uniform distribution of the magnetic
field along sample top and bottom faces. The exact calculation requires
knowledge of $A(x,y)$ inside the sample or ${\bf H}(x,y)$ in a screened
volume outside, proportional to $w^{2}$. The penetrated volume is:
\begin{equation}
V_p = \oint\limits_S {\frac{\lambda \left| {H_s } \right|}{H_0} ds}
\label{vp}
\end{equation}
\noindent where integration is conducted over the sample surface in a
$3D$ case or sample cross-section perimeter in a $2D$ case. Using
Eq.(\ref{hs}) for magnetic field on top and bottom surfaces and assuming
$H_s = H_0/(1-N)$ on sides we obtain:

\begin{equation}
-4 \pi \chi = \frac{1}{\left( 1-N \right)} \left[ 1 - \frac{\lambda}{R}
\tanh \left(\frac{R}{\lambda} \right) \right] \label{chifit}
\end{equation}
\noindent Here $N$ is an effective demagnetization factor and $R$ is the
effective dimension. Both depend on the dimensionality of the problem. As
mentioned earlier, Eq.(\ref{chifit}) is similar to the well-known
solution for the infinite slab of width $2w$ in parallel field. In that
case $R = w$ and the effective demagnetizing factor $N = 0$. In a 3D case
($2w \times 2w$ slab, infinite in the $z-$ direction), $R=w/2$ and $N=0$.
The $\tanh{(R/\lambda)}$ term in Eq. (\ref{chifit})was inserted to insure
a correct limit at $\lambda \rightarrow \infty$. This correction becomes
relevant at $\lambda/R \geq 0.4$, which is realized only at about
$T/T_{c} \geq 0.9$ for typical high-$T_{c}$ samples.
\begin{figure}
 \centerline{\psfig{figure=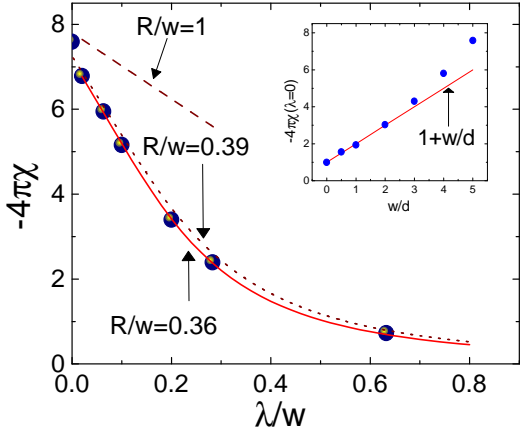,width=7cm,clip}}
 \caption{Calculated $-4\pi \chi \left( \lambda \right)$ for a slab of $w/d=5$.
 Solid line is a fit to Eq.(\ref{chifit}) with the effective
dimension $R/w=0.36$. Dotted line is calculated using $R/w=0.39$ from
Eq.(\ref{R2D}) and a dashed line is a plot with $R/w=1$. {\it Inset}: $-4
\pi \chi \left( \lambda \rightarrow 0 \right)$ calculated for samples of
different aspect ratio. Solid line is $1+w/d$.} \label{demag}
\end{figure}

For the actual geometry studied here, both $R$ and $N$ depend upon the
aspect ratio $w/d$. Unlike the case of an elliptical cross-section, the
magnetic field is not constant within the sample so there is no true
demagnetizing factor for a slab. However, $N$ can still be defined in the
limit of $\lambda \rightarrow 0$, through the relation, $4\pi
M/V_{s}=-H/\left( 1-N \right)$. We find numerically that in a 2D case, for
not too large aspect ratio $w/d$, $1/(1-N) \approx 1 + w/d$. Calculating
the expelled volume as described above, the effective dimension $R$ is
given by:
\begin{equation}
R_{2D}=\frac{w}{1+\arcsin{(a^{-1})}}
\label{R2D}
\end{equation}
\noindent In the thin limit, $ d \ll w$ ($a \rightarrow 1$), we obtain
$R_{2D} \approx 0.39 w$.

The natural extension of this approach for the 3D disk of radius $w$ and
thickness $2d$ leads to $1/(1-N) \approx 1 + w/2d$ and
\begin{equation}
R_{3D} = \frac{w}{2 \left( 1 + \left[ 1+( \frac{2 d}{w} )^{2} \right]
\arctan{ (\frac{w}{2 d})} - \frac{2 d}{w} \right)} \label{R3D}
\end{equation}
\noindent In a thin limit, $R_{3D} \approx 0.2 w$.  Eq.(6) was derived
for a disk but the more experimentally relevant geometry is a rectangular
slab. There is no analytical solution for the slab.  However, $a^{2} = 1
+ (2d/w)^{2}$ is relatively insensitive to ${w}$ in the thin limit and so
we approximate $w$ for a slab by the geometric mean of its two lateral
dimensions.  The validity of this approach will be determined shortly.

To verify Eqs.(\ref{chifit}) and (\ref{R2D}) we calculated $\chi \left(
\lambda \right)$ numerically. The result is shown it in Fig. \ref{demag}
by symbols. The solid line is a fit to Eq.(\ref{chifit}) with $N = 0.86$
and $R/w = 0.36$. The effective dimension calculated using Eq.
(\ref{R2D}) gives $R/w = 0.39$ and the corresponding susceptibility curve
is shown as a dotted line. The calculated effective demagnetization
factor is $N = 0.84$. It is seen that our approximations are reasonably
good. It should be borne in mind that these are all $2D$ results - the
sample extends to infinity in the z-direction. Demagnetizing effects are
significantly larger in two dimensions than in three owing to the much
slower decay of fields as we move away from the sample (compare $3D$
sphere, $N=1/3$, and cylinder in perpendicular field, $N=1/2$). Therefore
we expect our approximations to work better in three dimensions.
\begin{figure}
 \centerline{\psfig{figure=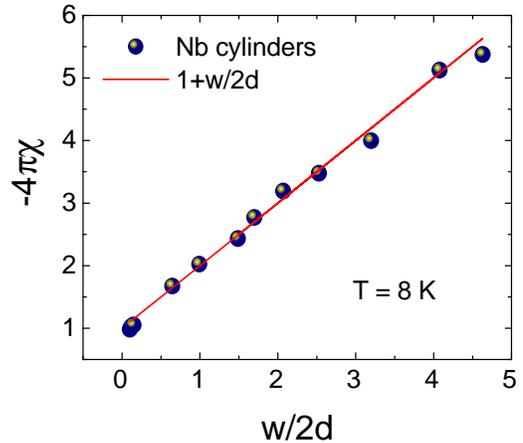,width=7cm,clip}}
 \caption{Linear magnetic susceptibility of Nb cylinders of different aspect
ratio measured at $T=8$ K. Solid line is a plot of $1+w/2d$}
 \label{nb}
\end{figure}

In a 3D case the validity of our results can be verified experimentally
by independently measuring the demagnetization factor as a function of
the aspect ratio and the magnetic susceptibility for a finite London
penetration depth $\lambda$. To achieve the first goal, we measured
niobium cylinders of radius $w$ and length $2d$ using a Quantum Design
MPMS-5 SQUID magnetometer. Sample dimensions were typically of the order
of millimeters, which allows us to disregard London penetration depth of
Nb (about 500 \AA). The initial susceptibility obtained from the
magnetization loops at $T=8$ K is shown in Fig. \ref{nb}. The solid line
is a plot of $1+w/2d$ (not the fit) and for an aspect ratio up to $w/d
=10$ the agreement is excellent.

To test our result for $R$ (Eq.\ref{R3D}) in actual samples
we need the magnetic penetration depth. It is common to measure changes in the
penetration depth by using the frequency shift of a microwave cavity or an
LC resonator. In these techniques, the relative frequency shift $\left(
f-f_{0}\right) /f_{0}$ due to a superconducting sample is proportional to
$H^{-2}\int {{\bf M}_{ac}\cdot {\bf H}}dV$, which in turn is proportional
to the sample linear magnetic susceptibility (${\bf M}_{ac}$ is the ac
component of the total magnetic moment, ${\bf H}$ is the external magnetic
field and $f_{0}$ is the resonance frequency in the absence of a sample).
Using Eq.(\ref{chifit}) and Eq.(\ref{R3D}) we obtain for $\lambda << R$:
\begin{equation}
\frac{\Delta f}{f_{0}}=\frac{V_{s}}{2V_{0} \left( 1- N \right)}\left( 1-
\frac{\lambda }{R}%
\right) \label{df}
\end{equation}
\noindent where $V_{s}$ is the sample volume, $V_{0}$ is the effective
coil volume. The apparatus and sample - dependent constant $\Delta f_{0}
\equiv V_{s}f_{0}/(2V_{0} \left( 1- N \right) )$ is measured directly
removing the sample from the coil. Thus, the change in $\lambda $ with
respect to its value at low temperature is

\begin{equation}
\Delta \lambda = -\delta f\frac{R}{\Delta f_{0}}
\label{dl}
\end{equation}
\noindent where $\Delta \lambda \equiv \lambda \left( T\right) -\lambda
\left( T_{\min }\right)$ and $\delta f \equiv \Delta f (T) - \Delta f
(T_{min})$.

\begin{figure}
 \centerline{\psfig{figure=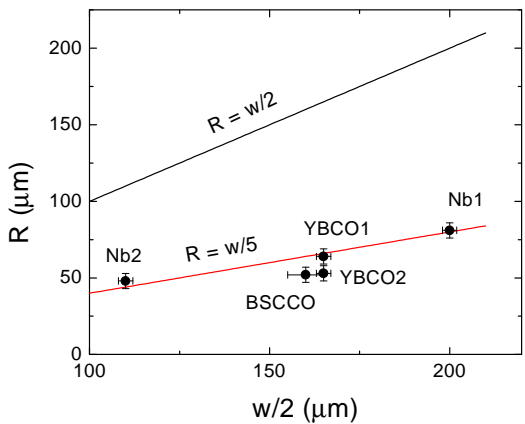,width=7cm,clip}}
 \caption{Effective dimension $R$ determined experimentally for different
 samples (symbols). The upper solid line is an "infinite slab" model ($R=w/2$)
 and the lower solid line is an analytic approximation $R \approx w/5$.}
 \label{rexp}
\end{figure}

We used an rf tunnel-diode resonator \cite{resonator} to measure $\delta
f$ in Nd foils, YBCO and BSCCO single crystals. Combining $\delta f$ with
an independent measurement of $\Delta \lambda (T)$ and a measured value
for $\Delta f_{0}$, we then arrived at an experimental determination of
the effective dimension $R$. For the Nb and YBCO samples, $\Delta \lambda
(T)$ was obtained using the demagnetization-free orientation (rf magnetic
field along the sample $ab-$ plane) where $R=w$ and $1/(1-N)=1$. In
BSCCO, the large anisotropy prohibits using this method and we used
reported values of $d\lambda /dT \simeq$ 10 \AA/K \cite{jacobs,slope}.
Figure Fig. \ref{rexp} summarizes our experimental results. The upper
line represents the "infinite slab" model, where $R=w/2$, whereas the
lower solid line is $R=0.2w$ obtained in a thin limit of Eq. (\ref{R3D}).
Symbols show the experimental data obtained on different samples,
indicated on plot. In three samples: YBCO1 (w/d = 57), Nb1 (w/d = 29) and
Nb2 (w/d = 15), $R$ agrees with Eq.(\ref{R3D}) to better than 5 \%. The
standard result, $R=w/2$, is too large by a factor of 2.5. Both YBCO2 and
BSCCO give $R$ roughly 20 \% smaller than predicted. For the BSCCO data,
it is possible that a sample tilt combined with the very large anisotropy
of $\lambda$ produces an additional contribution from $\lambda_{c}$. If
the c-axis is tilted by an angle, $\theta$ away from the field direction,
the frequency shift is given by
\begin{eqnarray}
\frac{\Delta f}{f_0} = \frac{V_{s}}{2V_{0} \left( 1- N \right)} \left( {1
- \frac{{\lambda _{ab} }}{R}} \right)\cos ^2 \left( \theta \right) +
\nonumber \\
\frac{V_{s}}{2V_{0}}\left( {1 - \left[ {\frac{{\lambda _{ab}
}}{d} + \frac{{\lambda _c }}{w}} \right]} \right)\sin ^2 \left( \theta
\right) \label{dftilt}
\end{eqnarray}

The importance of the tilt depends upon the relative changes in $\lambda
_{ab}$ and $\lambda _{c}$ with temperature. From Eq.(\ref{dftilt}) we
obtain for the relative contribution to the frequency shift:
\begin{equation}
 - \frac{{\delta f\left( \theta \right)}}{{\delta f\left( {\theta = 0} \right)}}
  \approx 1 + \frac{2}{5} \tan ^2 \left( \theta \right)  \left( {1 + \frac{d}{w}
 \frac{{\Delta \lambda _c }}{{\Delta \lambda _{ab} }}} \right)
 \label{deltaftilt}
\end{equation}

\noindent where we used the previous estimates of $N$ and $R$. For BSCCO
we take, $d\lambda _{c} /dT \simeq$ 170 \AA/K and $d\lambda _{ab} /dT
\simeq$ 10 \AA/K \cite{jacobs,slope}, Eq. (\ref{deltaftilt}) reduces to
$\approx 1+ \tan ^2 \left( \theta \right)$. We then find that for sample
tilt to produce an additional 20 \% frequency shift a misalignment of
$\theta \approx 20^{o}$ would be required. Our estimated misalignment was
a factor of 10 smaller than this so the discrepancy between measured and
predicted R was not due to tilt. Both the BSCCO and the YBCO2 sample were
more rectangular than square and our use of the geometric mean for $w$
could be the source of the error.

In conclusion, we solved numerically the London equations for samples of
rectangular cross-section in perpendicular magnetic field. We obtained
approximate formulae to estimate finite-$\lambda$ magnetic susceptibility
of platelet samples (typical shape of high-$T_{c}$ superconducting
crystals).

We thank M. V. Indenbom, E. H. Brandt, and J. R. Clem  for useful
discussions. This work was supported by Science and Technology Center for
Superconductivity Grant No. NSF-DMR 91-20000. FMAM gratefully
acknowledges Brazilian agencies FAPESP and CNPq for financial support.

\end{multicols}
\end{document}